\documentstyle{mn}
\begin{document}
\def\simlt{\mathrel{\rlap{\lower 3pt\hbox{$\sim$}}
        \raise 2.0pt\hbox{$<$}}}
\def\simgt{\mathrel{\rlap{\lower 3pt\hbox{$\sim$}}
        \raise 2.0pt\hbox{$>$}}}

\title[Constraints on the Clustering, Biasing and Redshift Distribution
of Radio Sources]
{Constraints on the Clustering, Biasing and Redshift Distribution
of Radio Sources}
\author[M. Magliocchetti, S.J. Maddox, O. Lahav, J.V. Wall]
{M. Magliocchetti $^1$,
S.J. Maddox $^1$, O. Lahav $^1$, J.V. Wall $^2$\\
$^1$Institute of Astronomy, Madingley Road, Cambridge CB3 0HA\\
$^2$Royal Greenwich Observatory, Madingley Road, Cambridge CB3 0EZ}

\maketitle
\begin{abstract}

We discuss how different theoretical predictions for the variance
$\sigma^2$ of the counts-in-cells distribution of radio sources can be
matched to measurements from the FIRST survey at different flux
limits.
The predictions are given by the integration of models for the angular
correlation function $w(\theta)$ for three different
functional forms of the redshift distribution $N(z)$, different
spatial correlation functions that match the observed present day
shape and by different evolutions of the bias $b(z)$ with redshift.
We also consider the two cases of open and flat Universes. 
Although the predicted $w(\theta)$ show substantial differences due
to differences in the $N(z)$'s, these differences are not significant
compared to the uncertainties in the current observations. 
It turns out that, independent of the geometry of the universe and the
flux limit, the best fit is provided by models with constant biasing
at all times, although the difference between models with
epoch-independent bias and
models with bias that evolves linearly with redshift is not very large.
All models with strong evolution of bias with epoch are ruled out
as they grossly overestimate the amplitude of the variance over the
whole range of angular scales sampled by the counts-in-cells analysis.
As a further step we directly calculated $w_{obs}(\theta)$ at 3mJy
from the catalogue and matched it with our models for the angular
correlation function in the hypothesis that the clustering signal comes
from two different populations, namely AGN-powered sources and
starbursting galaxies.
The results are consistent with a scenario for hierarchical clustering
where the fainter 
starbursting galaxies trace the mass at all epochs, while the
brighter AGN's are strongly biased, with $b(z)$ evolving
linearly with redshift, as suggested by some theories of galaxy formation
and evolution.

\end{abstract}

\section{INTRODUCTION}
During the past few years there has been an increasing interest in how
galaxies trace mass and specifically in the issue of a biasing factor
evolving with redshift (e.g. Fry, 1996; Bagla, 1997; Matarrese et al.,
1997; Moscardini et al., 1997; Tegmark \& Peebles, 1998). 
This has received added impetus after Steidel et al.  (1996, 1997)
reported evidence for a strong concentration of galaxies at $z\sim 3$
(Lyman-Break Galaxies), which would imply a value of $b\sim 6$ at
those redshifts.\\

\noindent     
Radio objects can be detected up to very significant cosmological
distances ($z\sim 4$) and therefore provide information on large scale
structure at rather early epochs when the main growth of
structures occurred.
Even though evidence of clustering in radio 
catalogues was detected in early wide-area surveys (Seldner \&
Peebles, 1981; Shaver \& Pierre,
1989; Kooiman et al., 1995; Sicotte, 1995; Loan, Wall \& Lahav, 1997),
the FIRST survey (Becker et al., 1995) is the first one in which 
the angular clustering
of radio sources down to the mJy level is detected with high
signal to noise ratio (Cress
et al., 1997; Magliocchetti et al., 1998). Unfortunately the relation 
between angular measurements and
spatial quantities is strongly dependent on
the radio source redshift distribution $N(z)$, 
which becomes more and more uncertain as
the flux threshold is lowered. Dunlop and Peacock (DP,
1990) provided models of epoch-dependent luminosity functions for
radio sources to make estimates of $N(z)$ that work well for
bright sources, but these predictions drastically diverge at faint
fluxes. One of the main uncertainties at such low flux densities is
given by the presence
and relative amplitude of a low-redshift
spike due to a population of starbursting
galaxies (e.g. Windhorst et al., 
1985) now believed to constitute a majority of sources at mJy levels 
(see Wall, 1994 for an overview). \\

\noindent
The knowledge of a well assessed functional form for the redshift
distribution of radio sources $N(z)$ has become of crucial importance in the
last decade for both radio-astronomy and cosmology. In the former case
this would provide tests for radio-source unification and luminosity
evolution models (Wall \& Jackson, 1997), while in the latter case 
it would allow conversion
of angular clustering measurements to spatial clustering estimates
that can be used to constrain structure formation models.\\

\noindent
We present here a theoretical approach able to put constraints on
both the functional form of $N(z)$ at fluxes $F\le 10$mJy and 
the clustering properties of radio objects. We pay particular
attention to the evolution of bias, i.e. the way radio galaxies trace
the mass distribution. 
We begin by assuming that the spatial correlation function of mass
is determined by the linear growth of mass fluctuations.  This is a
reasonable assumption because the mean redshift of radio sources in
the survey ($\bar{z}\sim 1$ as opposed to $\bar{z}\sim 0.1$ obtained
for optical and infrared surveys), and so the angular scales that we
consider correspond to physical scales where the linear theory still
holds.
The clustering of radio galaxies is then related to mass via 
three different models for the evolution of bias with redshift, each
one related to a reasonable model for galaxy evolution, as taken from
Matarrese et al., 1997 and Moscardini et al., 1997. 
We then predict the angular two-point correlation function $w(\theta)$
for three models of $N(z)$ that span the range realistic distributions
for faint radio objects.
We obtain the correlation function using both flat and open
geometry.\\

%The angular correlation function will 
%be obtained under the 
%assumption of growth of clustering under linear theory given that, 
%because of the
%mean redshift of radio sources in the survey ($\bar{z}\sim 1$ as
%opposed to $\bar{z}\sim 0.1$ obtained for optical and infrared
%surveys), information on clustering at those epochs is very likely to
%be related to
%scales where the linear regime still holds. The possibility of
%flat/open space is also taken into account.\\

\noindent
In section 2 we introduce the different models for $N(z)$ used in the
projection analysis, while section 3 describes the calculation of the
theoretical predictions for $w(\theta)$; the predictions 
for each model are then presented in section 4.
In section 5 we move to the analysis of the variance and we show the
results obtained for the measurements of the $\sigma^2$ of the
distribution of sources in the FIRST survey at flux limits up to 10
mJy.
The data are compared with our predictions from different models of
$w(\theta)$ in section 6.
Section 7 presents our results on the interpretation of clustering and
bias under the assumption that radio sources are formed by two
different populations. Section 8 summarises our conclusions.

\section{THE REDSHIFT DISTRIBUTION N(z)}
To test the stability of the predicted angular correlation function
given for different functional forms of the redshift distribution
$N(z)$, we use the Dunlop \& Peacock (DP, 1990) models of
epoch-dependent space density for radio sources. 
These models used Maximum Entropy analysis to determine
polynomial approximations to the luminosity
function and its evolution with redshift; the approach incorporated
the then-available identification and redshift data for complete
samples from radio surveys at several frequencies.\\

\begin{figure}
\vspace{8cm}  % amount of vertical space needed
\includegraphics{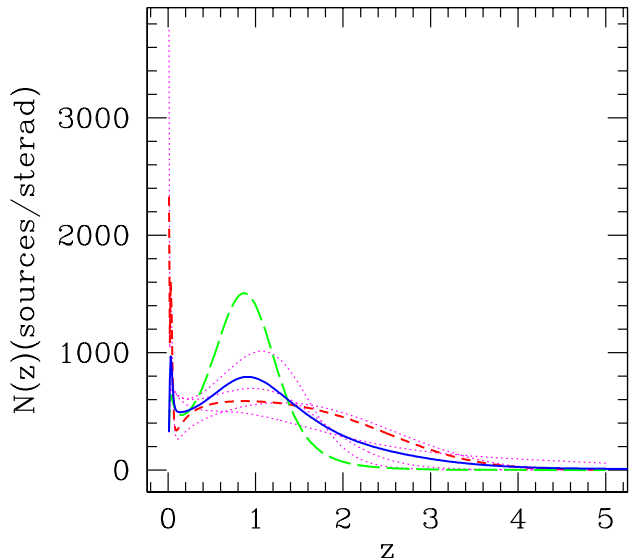} 
\caption{Redshift distribution $N(z)$ for the radio source population at
  1.4 GHz at a flux limit of 3 mJy. The dotted and dashed curves
  represent the 6
  models (1-4, 6-7) taken from Dunlop \& Peacock (1990); the solid curve
  is the average. Model 5 is omitted as it shows a totally unrealistic
  sharp and dominant feature at $z=4$
\label{fig:N_zgt3} }
\end{figure}

\begin{figure}
\vspace{8cm}  % amount of vertical space needed
\includegraphics{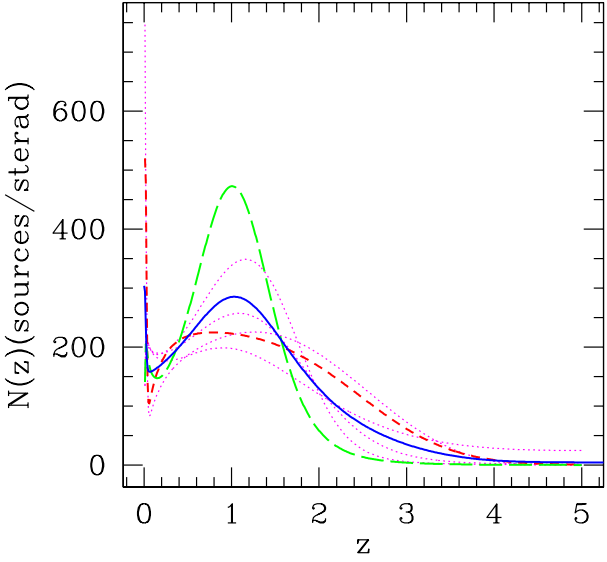} 
\caption{Redshift distribution $N(z)$ for the radio source population at
  1.4 GHz at 10 mJy. The dotted and dashed curves
  represent the 6
  models (1-4, 6-7) taken from Dunlop \& Peacock (1990); the solid curve
  is the average.
\label{fig:N_zgt10} }
\end{figure}

\noindent 
Figures \ref{fig:N_zgt3} and \ref{fig:N_zgt10} show, for flux limits
of 3 and 10 mJy respectively, models 1-4 and 6-7 taken from DP
(dotted-dashed lines) and their average (solid line).
The discrepancies amongst different models become increasingly
significant as the flux threshold is lowered, mirroring the lack of
any information on radio objects at faint fluxes.
We have chosen three models which span the range of reasonable 
$N(z)$; two representing the extremes, and one intermediate.\\

%Out of those models we have then chosen three  
%with completely different features in order to put upper and
%lower limits to the range of possibilities for N(z).

\noindent
The first model we adopted is DP's 7 (called hereafter N3), indicated 
by Peacock (private communication) as the best model, which predicts 
$N(z)$ for pure
luminosity evolution. This is illustrated in figures
\ref{fig:N_zgt3} and \ref{fig:N_zgt10} by the short-dashed line, and
is characterised by a 
low-redshift component which becomes more dominant at fainter fluxes, 
and by a broad and shallow maximum for $z\simgt0.5$. The low-z spike 
could be interpreted as modelling the starburst-galaxy population 
which appears at these faint fluxes.\\
The second model considered in our analysis, 
DP's 3, (called hereafter N2) is represented in figures
\ref{fig:N_zgt3} and \ref{fig:N_zgt10} by the long-dashed line; it has a
narrow and prominent bump around $z\sim 1$ and no low-redshift
spike.\\
As an intermediate model $N1$ we chose the average of DP's 
1-4, 6-7 for it shows both the
low-redshift component and the $z\sim 1$ bump, but neither feature
is too dominant. This is plotted as the solid line in Figures 1 and 2.  

%parlare di risultati di connolly et al.
\section{PREDICTING THE ANGULAR CORRELATION FUNCTION}
The standard way of relating the angular two-point correlation
function $w_{radio}(\theta)$ to the spatial two-point correlation function
$\xi_{radio}(r,z)$ is by means of the relativistic Limber equation (Peebles,
1980):
\begin{eqnarray}
w_{radio}(\theta)=2\:\frac{\int_0^{\infty}\int_0^{\infty}F^{-2}(x)x^4\Phi^2(x)\xi_{radio}(r,z)dx\:du}{\left[\int_0^{\infty}F^{-1}(x)x^2\Phi(x)dx\right]^2},
\label{eqn:limber} 
\end {eqnarray}
where $x$ is the comoving coordinate, $F(x)$ gives the correction for
curvature, and the selection function $\Phi(x)$ is determined by the
$N(z)$, 
\begin{eqnarray}
{\cal N}=\int_0^{\infty}\Phi(x) F^{-1}(x)x^2 dx=\frac{1}{\Omega_s}\int_0^{\infty}N(z)dz,
\label{eqn:Ndense} 
\end{eqnarray}
in which $\cal N$ is the mean surface density on a surface of solid angle
$\Omega_s$ and $N(z)$ is the number of objects in the given survey
within the shell ($z,z+dz$).\\ 
The physical separation between two sources separated by an angle
$\theta$ is given (in the small angle approximation) by:
\begin{eqnarray}
r\simeq\frac{1}{(1+z)}\;\left(\frac{u^2}{F^2}+x^2\theta^2\right)^{1/2}.
\label{eqn:r}
\end{eqnarray} 
\\

\noindent
The comoving coordinate
$x$ and the correction factor $F(x)$ are different for
different geometries; for a Universe with generic density parameter
$\Omega_0$ and cosmological constant $\Lambda=0$ (see e.g. Treyer \&
Lahav, 1996):
\begin{eqnarray}
x=\frac{2c}{H_0}\left[\frac{\Omega_0 z-(\Omega_0-2)(1-\sqrt{1+\Omega_0
      z})}{\Omega^2_0 (1+z)}\right]
\label{eqn:x}
\end{eqnarray}
and
\begin{eqnarray}
F(x)=\left[1+\left(\frac{H_0 x}{c}\right)^2(\Omega_0 -1)\right]^{1/2}.
\end {eqnarray}
This leads, with equation (\ref{eqn:Ndense}), 
to the following expression for the angular correlation function:
\begin{eqnarray}
w_{radio}(\theta)=2\frac{H_0}{c}\Omega^2_0\;\frac{\int_0^{\infty}
N^2(z)/P(\Omega_0,z) dz \int_0^{\infty}\xi_{radio}(r,z)du}
{\left[\int_0^{\infty}N(z)dz\right]^2},
\end{eqnarray}
where $P(\Omega_0,z)$ is given by
\begin{eqnarray}
P(\Omega_0,z)=\frac{4(\Omega_0-1)[(1+\Omega_0
  z)^{1/2}-1]+\Omega_0^2(1-z)+2\Omega_0 z}{(1+z)^2(1+\Omega_0 z)^{1/2}},
\end{eqnarray}
and $u$ is defined by (\ref{eqn:r}).\\ 

\noindent
In the case of a cosmological constant $\Lambda \ne 0$ with
$\Omega_0+\Lambda=1$ (flat space), we have $F(x)=1$ and:
\begin{eqnarray}
x=\frac{c}{H_0}\Omega_0^{-1/2}\int_0^z\frac{dz}{\left[(1+z)^3+
\Omega_0^{-1}-1\right]^{1/2}},
\end{eqnarray}
(see Peebles, 1984; Treyer \& Lahav, 1996) so that the expression for $w(\theta)$ assumes 
the form 
%(see e.g. Efstathiou et al., 1991):
 \begin{eqnarray}
w_{radio}(\theta)=2\frac{H_0}{c}\Omega_0^{1/2}\;\frac{\int_0^{\infty}N^2(z)
Q(\Omega_0,z)dz\int_0^{\infty}\xi_{radio}(r,z)
du}{\left[\int_0^{\infty} N(z) dz\right]^2}.
\end{eqnarray}
with
\begin{eqnarray}
Q(\Omega_0,z)=[(1+z)^3+\Omega_0^{-1}-1]^{1/2}
\end{eqnarray}
\\
\\
\begin{figure}%[htb]
\vspace{8cm}  % amount of vertical space needed
\includegraphics{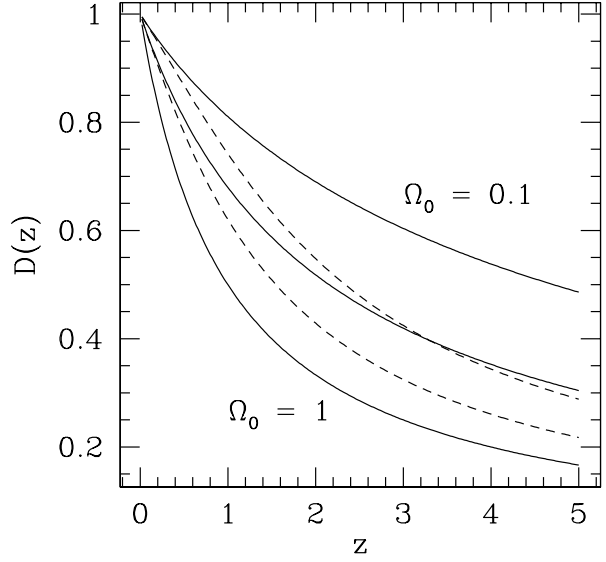} 
\caption{The linear-theory density growth $D(z)$ scaled to unity at
  the present time for $\Omega_0=1$, $\Omega_0=0.3$ and
  $\Omega_0=0.1$. The solid lines represent open models ($\Lambda=0$)
  while the dashed lines are for flat models ($\Lambda+\Omega_0=1$).
\label{fig:D_z} }
\end{figure}
\noindent
\begin{figure}%[htb]
\vspace{8cm}  % amount of vertical space needed
\includegraphics{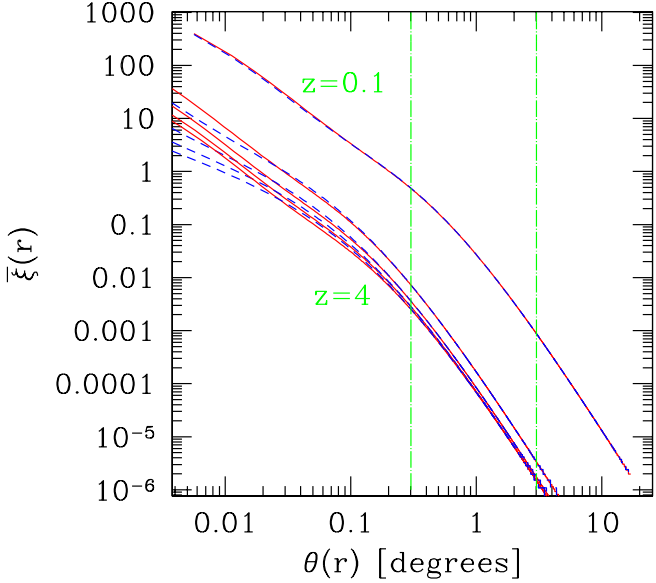} 
\caption{Integrated correlation function $\bar{\xi}(r,z)$ as a
  function of the angular scale $\theta$ in the case $\Omega_0=1$ at
  redshift $z=0.1, 1, 2, 3, 4$. Solid lines represent the predictions for
  growth of clustering under linear theory on the assumption of a
  $\xi_{mass}(r)$ as described in the text, while the dashed lines
  take into account the effects of non-linear evolution. The two
  vertical lines bracket the range $0.3^{\circ}\le\theta\le3^{\circ}$ 
provided by the observations.
\label{fig:evol} }
\end{figure}
\noindent
To take into account the evolution of clustering with epoch, 
we express the correlation function $\xi_{radio}(r,z)$ as:
\begin{eqnarray}
\xi_{radio}(r,z)=\xi_{mass}(r)D^2(z)b^2(z),
\label{eqn:x_i} 
\end{eqnarray}
We have chosen the present-day $\xi_{mass}(r)$ to have a shape which
matches the observed $\xi_{gal}$ from local optical galaxy surveys
(eg. APM, LCRS, etc.) and a normalization which matches the 4-year
COBE data (Bunn \& White 1997).  On large scales we use the
linear-theory prediction from the power spectrum $P(k)$ of Bond \&
Efstathiou (1984).  On small scales the linear prediction
under-estimates the true amplitude, and so we extrapolated with a
power-law of slope $-1.7$, as observed from the APM correlation
function (Maddox et al., 1990).  For $\Gamma = 0.2$ the extrapolation
is used for $r\le r_0^{gal}=5.4$Mpc. For other values of $\Gamma$ we
have re-scaled $\xi$ so that $\xi_1(r) \propto \xi_2(r \Gamma1 /
\Gamma2 )$.  In each case we use the 4 year COBE data to fix the
normalization in terms of mass fluctuations $\sigma_8^{mass}$.

\noindent
$D(z)$ is the linear-theory density growth-rate whose 
generic expression is given by Peebles (1980) and Lahav et al. (1991); 
$D(z)=(1+z)^{-1}$ for an
Einstein-de Sitter Universe
\begin{eqnarray}
f(z)\equiv\frac{d\;\ln(D(z))}{d\;\ln(a)}
\label{eqn:D_z}
\end{eqnarray}
with $a=(1+z)^{-1}$ and $f(z)\simeq R^{0.6}$, where
\begin{eqnarray}
R(z,\Omega_0,\Lambda)=\Omega_0(1+z)^3
[\Omega_0(1+z)^3- \nonumber \\
(\Omega_0+\Lambda-1)(1+z)^2+\Lambda]^{-1}
\end{eqnarray}

\noindent
Figure \ref{fig:D_z} shows the behaviour of $D(z)$ for a range of 
$\Omega_0$ and $\Lambda$. Our choice to consider only the case for
growth of clustering under linear theory 
is extensively discussed in Magliocchetti et al., 1998; 
given the high median redshift of
objects in the FIRST survey ($\bar{z} \sim 1$), the data obtained in the
angular range $0.3^{\circ}\le\theta\le 3^{\circ}$ provided by our 
analysis of the catalogue correspond to
a range of spatial scales ($r>10h^{-1}$Mpc) where the linear regime
still holds.  
As a check we calculated the
integrated correlation function $\bar{\xi}(r,z)\equiv
\frac{3}{r^3}\int_0^r \xi(y,z)\;y^2\;dy$ as obtained for
$\xi(r,z)=\xi(r)D^2(z)$ (growth of clustering under linear theory) 
in the case of
$\xi_{mass}(r)$ as described earlier in this section and
$b(z)=1$. This integral was performed at different redshifts and
for $\Omega=1$ and the resulting integrated correlation functions have 
been compared with the expression 
\begin{eqnarray}
\bar{\xi}=\frac{x+0.358x^3+0.0236x^6}{1+0.0134x^3+0.00202x^{9/2}}\;\;\;\;
x=a^2\bar{\xi}_0(r_0)
\label{eqn:evol}
\end{eqnarray}
from Hamilton et al., 1991 to take into account the
effects of non-linear evolution. The results are shown in figure
\ref{fig:evol} where we plotted $\bar{\xi}(r,z)$ as a function of the
angular scale $\theta=r/xa$ with $x$ given by equation \ref{eqn:x} in
the case $\Omega_0=1$, for $z=0.1, 1, 2, 3, 4$; solid lines represent the
evolution of the integrated correlation function under linear theory,
while the dashed lines are obtained from equation (\ref{eqn:evol}). It
is clear from the plot that differences between the two models occur
only at angular scales significantly smaller than the range of our
observations ($0.3^{\circ}\le\theta\le3^{\circ}$) which are bracketed
by the two dashed vertical lines. This gives some confidence that our
model assumptions are reasonable.\\
%
%
%As
%it is clear from the plot, differences between the two models only
%occur at angular scales sensibly smaller than our range of
%observations ($0.3^{\circ}\le\theta\le3^{\circ}$) bracket by the two
%vertical lines, therefore proving once again the validity of our assumptions.

\begin{figure}%[htb]
\vspace{8cm}  % amount of vertical space needed
\includegraphics{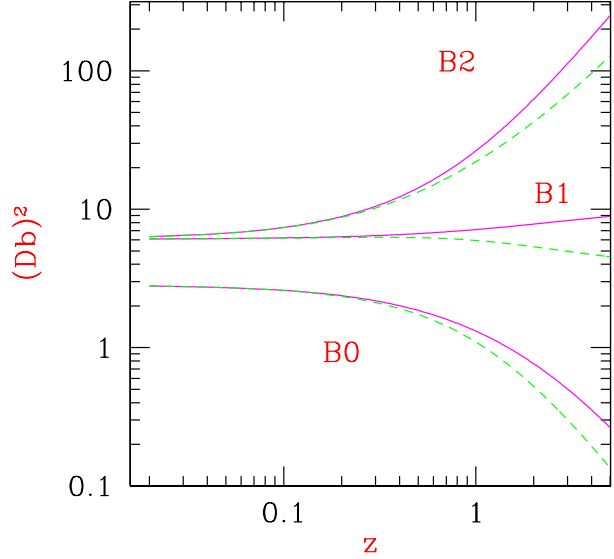} 
\caption{Trend of the quantity $D(z)\;b(z)$ as a function of 
  redshift for three different models of biasing as described in the
  text. $\Omega_0$ is set to be 0.3, while $b_0^{opt}=1.46$. 
The solid lines represent open models
  while the dashed lines are for flat models.  
\label{fig:D2_z} }
\end{figure}

\noindent
The biasing factor $b(z)$ in equation (\ref{eqn:x_i}) allows for
evolution in the way radio sources trace the mass distribution. 
In principle $b$ can be a function of both scale and redshift, but at
$z=0$, significant deviations from $b(r)=const$ are seen only on small
scales ($r< 10h^{-1}$Mpc) (Bagla, 1997). Thus in the following analysis
we ignore any variation of bias with scale.\\
\noindent
We take into consideration three possible models for the
evolution of biasing as a function of the redshift (for an extensive
study see Matarrese et al., 1997 and Moscardini et al., 1997).  
The first of these, called hereafter B0, is:
\begin{eqnarray}
b(z) \equiv b_0\simeq \frac{1}{\sigma_8^{mass}}\;\;\left(\frac{r_0}{5.4}\right)^{1.7/2},
\end{eqnarray}
representing constant bias at all epochs. The two other models are derived
from the expression
\begin{eqnarray}
b(z)=b_{-1}+(b_0-b_{-1})(1+z)^{\beta}
\label{eqn:b_z}
\end{eqnarray}
with 
\begin{eqnarray}
b_0\simeq
\frac{1}{\sigma_8^{mass}}\;\;\left(\frac{r_0}{5.4}\right)^{1.7/2}.\nonumber
\end{eqnarray}
Setting $b_{-1}=\beta=1$ gives model B1, while setting  
$b_{-1}=0.41$, $\beta\simeq 1.8$ gives model B2.\\
The B1 parameters have been chosen in agreement with the
so-called {\it galaxy conserving model} (Fry, 1996; Tyson, 1988) in
which galaxies form at some particular redshift and then evolve
without losing their identity, by following the continuity
equation.
The latter set (B2) refers to results from N-body simulations (see e.g. Bagla,
1997) and describes the so-called {\it merging model} in which faint
galaxies are subunits that merge to make up more luminous galaxies
(Broadhurst et al., 1992; Clements \& Couch, 1996; Baugh et al.,
1996). $r_0$
is the clustering length of radio sources at $z=0$
($\sim10h^{-1}$Mpc in our case, as measured by Peacock \& Nicholson,
1991; see also Magliocchetti et al., 1998) while
$r_0^{gal}=5.4h^{-1}$Mpc is the
corresponding value for optically-selected galaxies 
as measured in the APM survey, 
so that the factor $\left(\frac{r_0}{5.4}\right)^{1.7/2}$ allows for 
the different bias level of 'optical' galaxies
compared to radio objects as seen at zero redshift.\\
The mass r.m.s. fluctuation amplitude inside a sphere of $8h^{-1}$Mpc is
given by $\sigma_8^{mass}$, and the quantity
$b_0^{opt}=1/\sigma_8^{mass}$ 
is the
bias of the distribution of optical galaxies relative to the
distribution of mass. This value depends on both the geometry and the
normalisation of the power spectrum $P(k)$.  Note that the amplitude
of $\xi_{mass}(r)$ in equation (11) depends on the evolution of the
biasing factor with redshift given in equations (15-16) through the
relation:
\begin{eqnarray}
\xi_{radio}(r,0)=b^2(0)\xi_{mass}(r).
\end{eqnarray}
From equation (17) we can therefore determine the values for the mass
correlation length $r_0^{mass}$ that turns out to be:
%\begin{eqnarray}
%r_0^{mass}=5.4 h^{-1}\; \mbox{Mpc} \nonumber
%\end{eqnarray}
%for the B0 models, and
\begin{eqnarray}
r_0^{mass}=8\;\;\left(\frac{(\sigma_8^{mass})^2}{c_{\gamma}}\right)^{1/\gamma},
\end{eqnarray}
where $c_{\gamma}$ is a factor 
depending on the slope of the correlation function at small scales
(see Peebles, 1980).

\noindent
As an illustration figure \ref{fig:D2_z} shows the quantity
$D(z)\;b(z)$ as a function of redshift for the three biasing models
described by equations (15-16) for $\Omega_0=0.3$, flat (dashed
lines)/open (solid lines) Universe and $b_0^{opt}=1.46$.
The evolution of clustering with redshift/epoch is completely
determined by the combination of bias $b(z)$ and density growth rate
$D(z)$ (equation \ref{eqn:x_i}), and so the different forms of $b(z)$,
give models where a) clustering decreasing with look-back time ($B0$),
b) clustering roughly constant ($B1$) and c) clustering rapidly
increasing with look-back time ($B2$). \\

\noindent
To constrain an already wide parameter space we fix the
value for the reduced Hubble constant as $h_0=0.65$ (see
e.g. Kundic et al., 1997). 
%Furthermore,
%to check the dependence of the predicted angular correlation function
%on the geometry of the Universe we
%have plotted in figure \ref{fig:omegadep} the CDM predictions for 
%$w(\theta)$ in the case of $\Gamma=0.2$, N(z) given by N1 at 3mJy,
%two different biasing models, namely $b^2(z)=const=\left(\frac{r_0}{5.4}\right)^{1.7}$ (B0 - lower
%curves) and b(z)
%described by B2 (upper curves), in both the cases of open and flat
%Universe and different values of the density parameter 
%$\Omega_0$. 
%Moreover, in the framework of this analysis it is almost impossible to
%discriminate between different values of $\Omega_0$, and so we
%restrict our analysis to $\Omega_0=0.4$, allowing for flat geometries
%($\Omega_0+\Lambda=1$).
%This degeneracy in $\Omega_0$ is
%referred to as {\it cosmic  conspiracy}: low-density models in which
%dynamics yields less rapid clustering evolution are also those in
%which geometry provides more time for evolution.\\
Moreover, it is almost impossible to discriminate between different
values of $\Omega_0$ through the evolution of clustering, because low
density models in which clustering evolves slowly have a geometry which
provides more time for evolution. This degeneracy in $\Omega_0$ is
often referred to as {\it cosmic conspiracy}. Hence we restrict our
analysis to $\Omega_0=0.4$, allowing also for flat geometries
($\Omega_0+\Lambda=1$). 
We adopt the normalisation of the r.m.s. mass fluctuations,
$\sigma_8^{mass}$, determined from the 4-year COBE 
data by Bunn \& White (1997). 
So choosing a well-defined cosmology (i.e. a fixed value of $\Omega_0$
and $\Lambda$) fixes the value of $\sigma_8$ and therefore the
amplitude of $\xi_{mass}(r)$, given by $r_0^{mass}$ from equation (18)
Then the parameter $b_0^{opt}$ in the expression for $b(z)$ is also
determined by equation (15) and (16). We then have:
\begin{eqnarray}
\Omega_0=0.4\;\;\Lambda=0\;\;\rightarrow
\sigma_8^{mass}=0.64;\;
r_0^{mass}=3.31h^{-1}\mbox{Mpc}\nonumber\\
 \Omega_0=0.4\;\;\Lambda=0.6\rightarrow
\sigma_8^{mass}=1.07;\;\;
r_0^{mass}=6.23h^{-1}\mbox{Mpc}\nonumber  
\end{eqnarray}
As a comparison for a standard CDM model we would have
$\sigma_8^{mass}=1.22;\;\;r_0^{mass}=6.82h^{-1}$Mpc. Thus 
it turns out that in $\Lambda$CDM and
standard CDM (SCDM) models the distribution of optical galaxies is
``antibiased'' relative to the distribution of mass so that the
amplitude of the corresponding $\xi^{mass}(r)$ is higher than that
found for galaxies; the opposite case pertains for open CDM
(OCDM) models.\\
Concerning the constant bias case B0 we have instead decided to fix
(ad hoc) $\sigma_8^{mass}=1$, so that $r_0^{mass}=5.4$ h$^{-1}$Mpc.\\

\noindent
  Note that in the case of $\Gamma=0.5$ both $r_0^{mass}$ and
  $\sigma_8^{mass}$ will differ from the values expressed for $\Gamma=0.2$. In
  more detail we have:
\begin{eqnarray}
\Omega_0=0.4\;\;\Lambda=0\;\;\rightarrow
\sigma_8^{mass}=1.51;\;\;
r_0^{mass}=8.95h^{-1}\mbox{Mpc}\nonumber\\
 \Omega_0=0.4\;\;\Lambda=0.6\rightarrow
\sigma_8^{mass}=2.52;\;\;
r_0^{mass}=15.8h^{-1}\mbox{Mpc},\nonumber  
\end{eqnarray}
i.e. if $\Gamma=0.5$, in both the $\Lambda$CDM and 
  OCDM models the distribution of galaxies is ``antibiased'' relative
  to the distribution of mass.\\

\section{PREDICTIONS FOR $w(\theta)$}
%\begin{figure}
%\vspace{10cm}  % amount of vertical space needed
%\special{psfile=fitdiffomega.ps hoffset=-40 voffset=-55 vscale=50 hscale=50} 
%\caption{ CDM prediction for the angular correlation function
%  $w(\theta)$ in the case of 
%$\Gamma=0.2$, N(z) given by N1 at 3mJy and different values of
%$\Omega_0$. The solid lines
%correspond to open models in the constant bias case, the long dashed lines
%correspond to spatially flat models ($\Lambda\ne 0$) with
%$b^2=const=\left(\frac{r_0}{5.4}\right)^{1.7}$, the
%short dashed lines refer to open models with biasing evolution
%described by B2 (see section 3) while the dotted lines are for flat
%models and B2.
%\label{fig:omegadep} }
%\end{figure}
In this section we will focus our attention on the predicted
angular correlation function $w(\theta)$ obtained for two different CDM
models with respectively, $\Gamma=0.5$
(standard CDM model) and $\Gamma=0.2$ (modified CDM model), values that
bracket the range of models providing the best fit to the present 
observations of the power spectrum $P(k)$ (Peacock \& Dodds,
1994). Note that these models treat $\Gamma$ as a free parameter
independent of $h_0\Omega_0$, and so do not represent consistent CDM
models.\\
 \begin{figure*}
\vspace{15cm}  % amount of vertical space needed
\includegraphics{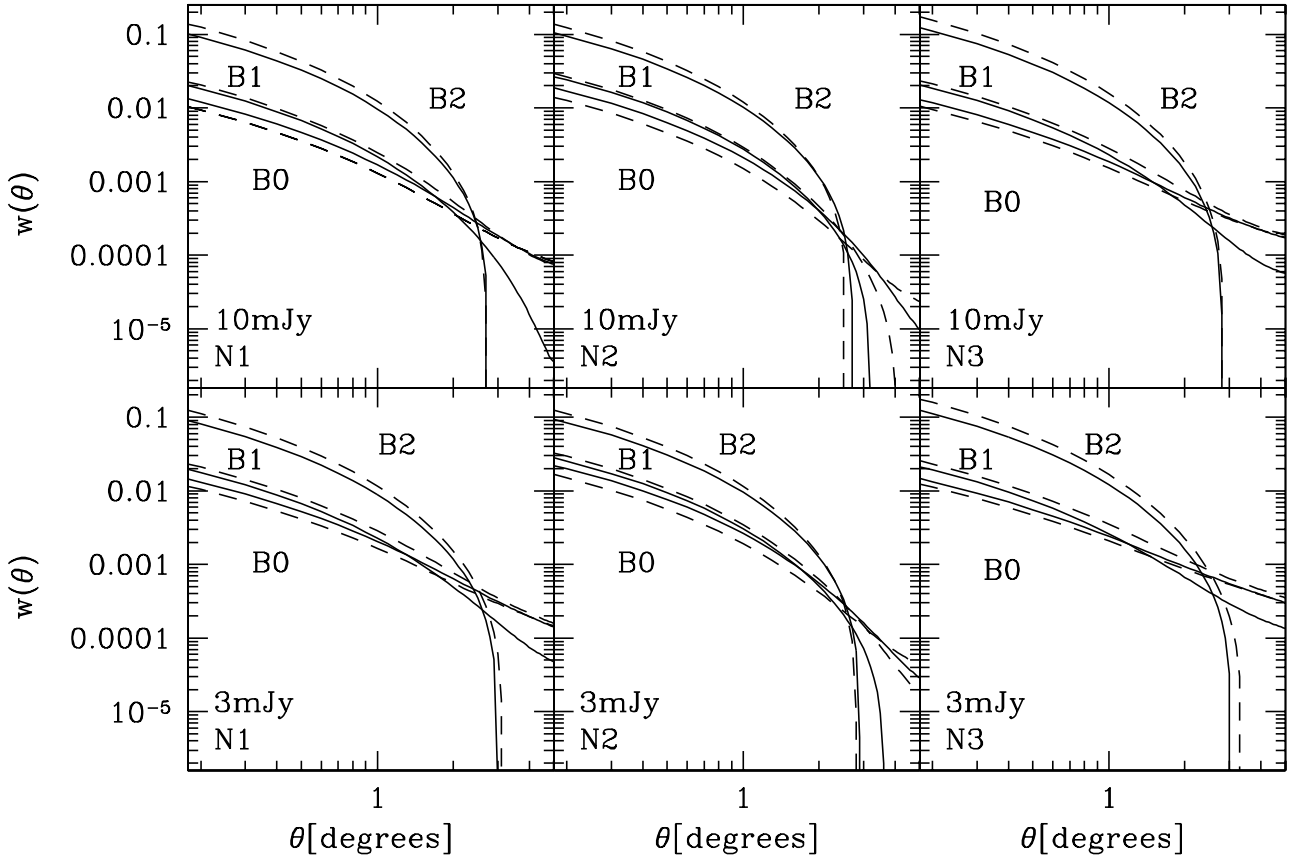} 
\caption{Theoretical prediction of the angular correlation function for
  $\Gamma=0.2$, $N(z)\equiv N1,N2,N3$ at 10mJy and 3mJy, 
  and different bias models as described in the text. 
The solid lines are for open models while the
  dashed lines indicate spatially-flat models.  
\label{fig:fit1} }
\end{figure*}
     
\noindent
We will focus on the three different redshift distributions of radio
sources (N1, N2, N3) described in section 2, and on the three
functional forms for the evolution of the biasing with redshift (B0,
B1, B2) introduced in section 3.  In order to test the
self-consistency of the predictions of each model, we consider two
different flux-cuts: 3mJy, the lowest flux limit where the
incompleteness of the survey is negligible (see Magliocchetti et al.,
1998) - and 10mJy, the brightest flux limit for which the measurements
of $w_{obs}(\theta)$ are not dominated by the errors).
In what follows we always start by considering models derived for
$F>10$mJy, because the differences between the different $N(z)$'s are
less significant than for fainter fluxes (see figures
\ref{fig:N_zgt3} and \ref{fig:N_zgt10}). In particular the relative
weight of the low-redshift component and of the $z\sim 1$ bump are 
much less variable at higher flux limits.\\
%
%We will focus on three different, and somehow ``extreme'' models
%for the redshift distribution of radio sources (N1, N2, N3)
%as described in section
%3, and on the three functional forms for the evolution of the
%biasing with redshift (B0, B1, B2) introduced in section 4. The analysis
%will be also carried out for two different flux-cuts (i.e. 3mJy -
%lowest limit for the flux given the incompleteness of the survey (see
%Magliocchetti et al., 1998) - and
%10mJy that is the brightest attainable value for which the measurements of
%$w_{obs}(\theta)$ are not dominated by the errors) in order to test the
%self-consistency of the predictions of each model. In what follows 
%we will always start by considering models derived for $F\ge 10$mJy as
%the difference amongst different N(z)'s is less striking than in the
%fainter fluxes case (see figures \ref{fig:N_zgt3} and
%\ref{fig:N_zgt10}), especially for what concerns the relative weight
%of the low-redshift component.\\ 
\begin{figure*}
\vspace{15cm}  % amount of vertical space needed
\includegraphics{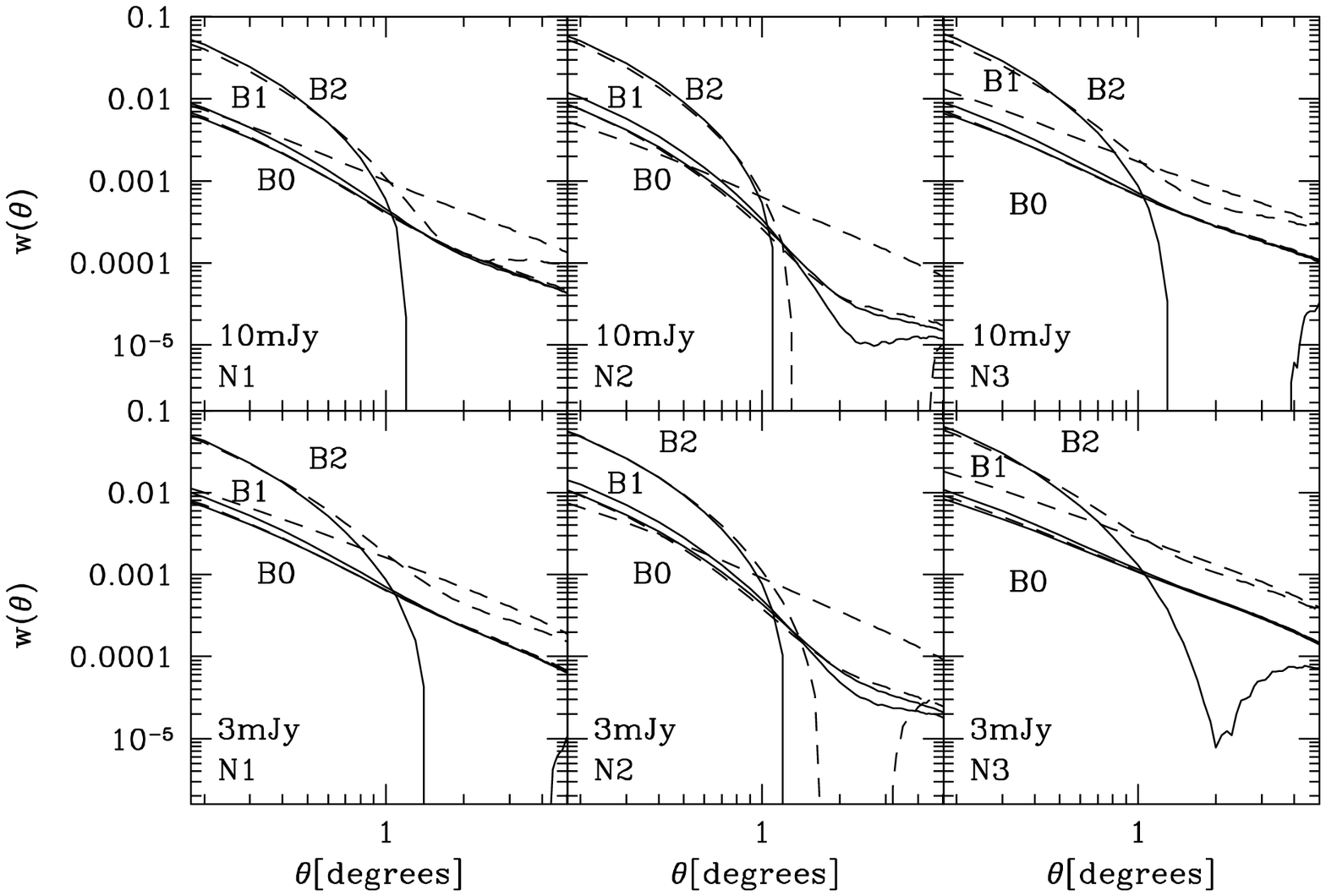} 
\caption{Theoretical prediction of the angular correlation function for
  $\Gamma=0.5$,  $N(z)\equiv N1,N2,N3$ at 10mJy and 3mJy, 
  and different bias models as
  shown in the figure. The solid lines are for open models while the
  dashed lines indicate spatially flat-models. 
\label{fig:fit11} }
\end{figure*}

\noindent
We will first show the results for a CDM power spectrum with
$\Gamma=0.2$ and then compare them to those with $\Gamma=0.5$. 
The theoretical estimates of $w(\theta)$ obtained at 
10mJy are to be compared with those 
for 3mJy, while keeping all the other
parameters fixed, in order to test the consistency of the predictions
of different models for $N(z)$.\\

\noindent
In Figure 6 we plot the theoretical estimates of $w(\theta)$ (in
the case $\Gamma=0.2$) at 10mJy (upper panel) and 3mJy (lower panel),
for the three models of $N(z)$ introduced in section 2 and the three
models for the evolution of bias with redshift of section
3. The correlation length $r_0$ in equations 15 and 16 has been fixed at
10$h^{-1}$ Mpc in agreement with both our previous estimate (Magliocchetti
et al., 1998) and the results found by Peacock and Nicholson (1991). The
solid lines are for open ($\Omega=0.4$) models while the dashed lines
are for flat ($\Lambda\ne 0$) geometries.\\

\noindent
Independent of the flux cut, 
the discrimination amongst different models comes both from
the eventual fall of $w(\theta)$ at large angular scales and from its
overall amplitude. The overall amplitude of the angular correlation
function is mainly related to the level of bias at high $z$: models
characterised by a bias factor increasing with redshift (B1/B2) 
show a stronger
clustering signal, especially at smaller angular scales, relative to
models with constant bias (B0). Their amplitude increases as
the dependence of $b(z)$ on the factor $(1+z)$ in equation (16)
becomes steeper, as can be seen by comparing the B2 with the B1
models. The difference in amplitude between the B0 and B1 models is not
pronounced; this effect is primarily due to the lower normalisation
of the B1 models deriving from having taken $\sigma_8^{mass}\ne 1$ as opposed
to the assumption of $\sigma_8^{mass}=1$ in the B0 case. Furthermore, 
as the main contribution to the clustering signal at large angular
  scales is given by low-redshift objects (see Cress \& Kamionkowsky,
  1998), the factor ($1+z$) is not strong enough to drive
  the evolution of the B1 models far from that described by B0.\\
The fall at large
angular scales depends instead on the negative feature in $\xi$, 
a genuine feature of CDM-like models, as well as on the amplitude of
the low-z component in the
$N(z)$. Note that in equations 6 and 9 the integral is weighted by
$N(z)^2$, so the spike makes a significant contribution to $w(\theta)$
even though the volume is small. The low-z amplitude describes 
the relative contribution of the local population 
of radio objects, that, as already mentioned, is the component which 
provides the dominant contribution to
the clustering signal at large angular scales (Cress \& Kamionkowsky,
1998). 
In fact from figure 6 one can see how all the correlation functions
obtained for N2 (no low-redshift component) drop at $\theta\sim
2^{\circ}$, while those for N3 and N1 models, characterised by a
larger low-redshift component, flatten out at this angular
scale, with a positive tail at larger scales that becomes more
predominant as the number of nearby objects is increased.
This effect is stronger at 3mJy than at 10mJy because the amplitude 
of the local population of star-bursting galaxies increases as the
flux threshold is lowered.\\

\noindent
The predictions for $w(\theta)$ for the case of $\Gamma=0.5$
are illustrated in figure 7; the models show basically the same
features as those obtained in the analysis for $\Gamma=0.2$, 
indicating
very little dependence on the value of the shape parameter $\Gamma$.\\
%
%plugged in the calculations of $\xi_{radio}(r)$ in equation (11).\\
\section{RESULTS FROM THE COUNTS IN CELLS ANALYSIS AT DIFFERENT FLUX LIMITS}
This section considers the counts-in-cells analysis of
the FIRST catalogue carried out at different flux limits; we 
then compare the predictions with the corresponding
measurements in section 6.
\begin{figure*}
\vspace{10cm}  % amount of vertical space needed
\includegraphics{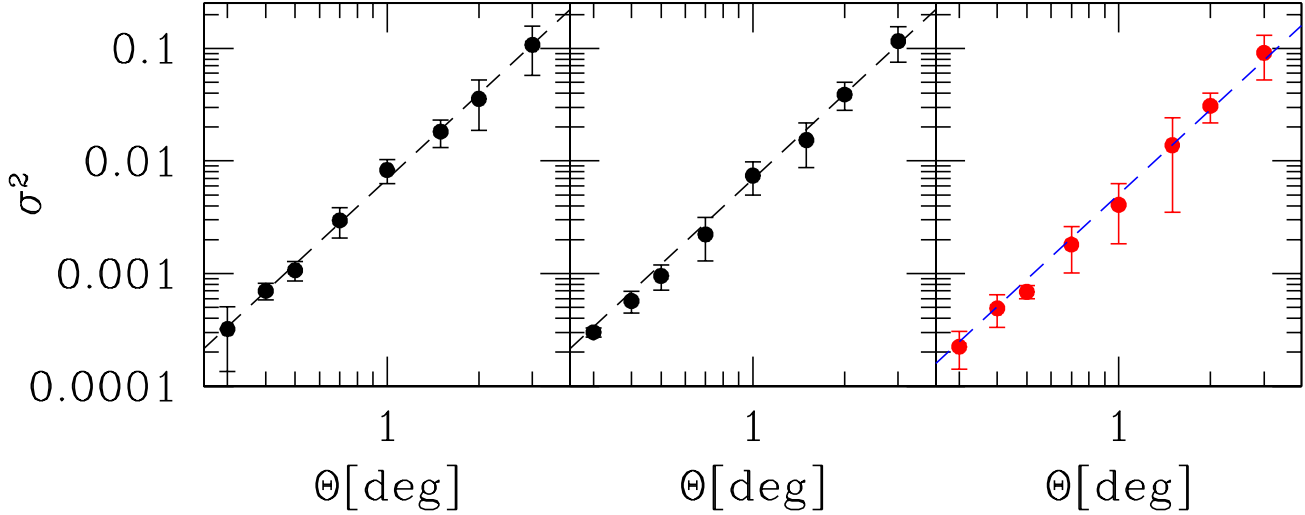}
%\vspace{8cm}  % amount of vertical space needed
%\special{psfile=figure8.ps hoffset=-40 voffset=-85 vscale=45 hscale=45} 
\caption{The normalised variance $\sigma^2$ vs the cell size $\Theta$
  for the FIRST survey for objects brighter than 5mJy (left panel),
  7mJy (middle panel) and 10mJy (right panel). 
Errors are estimated from the variance in four random subsets.
\label{fig:sigma5} }
\end{figure*}
%\begin{figure}%[htb]
%\vspace{8cm}  % amount of vertical space needed
%\special{psfile=figure9.ps hoffset=-40 voffset=-85 vscale=45 hscale=45} 
%\caption{The normalised variance $\sigma^2$ vs the cell size $\Theta$
%  for the FIRST survey for objects brighter than 7mJy. Errors are estimated from the variance in four
%  random subsets.
%\label{fig:sigma7} }
%\end{figure}
%\begin{figure}%[htb]
%\vspace{8cm}  % amount of vertical space needed
%\special{psfile=figure10.ps hoffset=-40 voffset=-85 vscale=45 hscale=45} 
%\caption{The normalised variance $\sigma^2$ vs the cell size $\Theta$
%  for the FIRST survey for objects brighter than 10mJy. Errors are estimated from the variance in four
%  random subsets.
%\label{fig:sigma10} }
%\end{figure}
\begin{figure}%[htb]
\vspace{8cm}  % amount of vertical space needed
\includegraphics{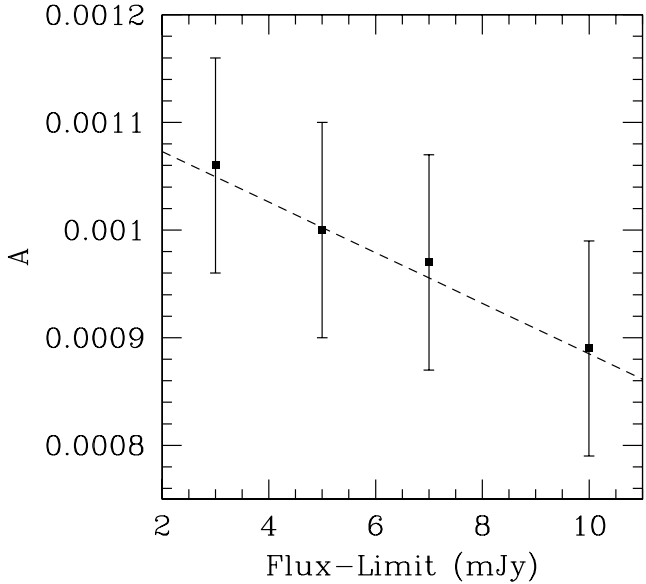} 
\caption{Amplitude of the angular correlation function
  $w_{obs}(\theta)$ as measured from the FIRST survey as a function of
  the flux limit.
\label{fig:ampl} }
\end{figure}
We measured the clustering of
radio objects using the so-called {\it galaxy distribution function} 
that gives the probability of finding $N$ objects in a
cell of particular size and shape. By defining the $k$-th moment of
the counts as
\begin{eqnarray}
\mu_k=\left<(N-\bar{N})^k\right>
\label{eqn:mu_k}
\end{eqnarray}
where $\bar{N}=n \Omega$ is the mean count in the solid angle
$\Omega$, we have that the second moment $\mu_2$ of the galaxy 
distribution function
is related to the two-point correlation function $w_{12}$ through the
expression
\begin{eqnarray}
\mu_2=\bar{N}+(\bar{N})^2\Psi_2
\label{eqn:mu_2}
\end{eqnarray}
where $\bar{N}$ is the shot noise resulting from the discrete nature
of the sources (Poisson noise), and
\begin{eqnarray}
\Psi_2\equiv\frac{1}{\Omega^2}\int w_{12}\:d\Omega_1\:d\Omega_2
\label{eqn:Psi_2}
\end{eqnarray}
is the normalised variance in terms of the two-point correlation
function integrated over a cell of area $\Omega$ 
and particular shape. By assuming a power-law form 
\begin{eqnarray}
w(\theta)=A\;\theta^{1-\gamma},
\label{eqn:w_power}
\end{eqnarray}
and by considering square cells of size $\Omega=\Theta\times\Theta$
square degrees, we can evaluate the integral in equation (\ref{eqn:Psi_2}) (see
Totsuji \& Kihara, 1969), obtaining
\begin{eqnarray}
\sigma^{2}\equiv\frac{\mu_2-\bar{N}}{(\bar{N}/\Omega)^2}=\int
A\:\theta^{1-\gamma}d\Omega_1d\Omega_2=
A\:C_{\gamma}\Theta^{5-\gamma}
\label{eqn:sigma2}
\end{eqnarray}
where $C_{\gamma}(\gamma)$ is a coefficient which can be
evaluated numerically by Monte Carlo methods (Lahav \& Saslaw,
1992). It is therefore possible to use the 
$\sigma^2 - \Theta$ relation to evaluate the
two parameters ($A$, $\gamma$) which describe the correlation
function (22).\\

\noindent
The following analysis has been carried out as in Magliocchetti et
al. (1998) for different versions of the original FIRST catalogue obtained by
combining source components into single source,
following well defined criteria. 
The results are shown in figure \ref{fig:sigma5} where we plot
$\sigma^{2}$ as a function of $\Theta$ for
different flux limits. The slopes and the intercepts of the plots are 
estimated by
a least-squares procedure minimising the quantity
\begin{eqnarray}
 \chi^2(a,b)=\sum_{i=1}^{N}\left(\frac{\log(\sigma)_i-a-b\log(\Theta)_i}{\Delta_i}\right)^2
\label{eqn:chi2}
\end{eqnarray}
with $b\equiv(5-\gamma)$, $a\equiv
\log(AC_{\gamma})$. The errors $\Delta_i$ are
obtained using the `partition bootstrap method' in which the
normalised variance is calculated for four subdivisions of the survey
region and the standard deviation of these measurements at each angle
is used as a measure of the error. From this analysis we 
find, for $F\ge3$mJy, $F\ge5$mJy, $F\ge7$mJy and $F\ge10$mJy respectively: 
\begin{eqnarray}
\gamma=2.50\pm 0.1 \;\;\; ; \;\;\; A=(1.06\pm 0.1)\cdot 10^{-3},
\label{eqn:sig2_fit3}
\end {eqnarray}
\begin{eqnarray}
\gamma=2.50\pm 0.1 \;\;\; ; \;\;\; A=(1.00\pm 0.1)\cdot 10^{-3},
\label{eqn:sig2_fit5}
\end {eqnarray}
\begin{eqnarray}
\gamma=2.50\pm 0.09 \;\;\; ; \;\;\; A=(0.97\pm 0.1)\cdot 10^{-3},
\label{eqn:sig2_fit7}
\end {eqnarray}
\begin{eqnarray}
\gamma=2.40\pm 0.2 \;\;\; ; \;\;\; A=(0.89\pm 0.1)\cdot 10^{-3},
\label{eqn:sig2_fit10}
\end {eqnarray}
The 3mJy measurement is taken from Magliocchetti et al. (1998).  Above
10mJy the measurements are dominated by Poisson errors.
 \begin{figure*}
\vspace{14cm}  % amount of vertical space needed
\includegraphics{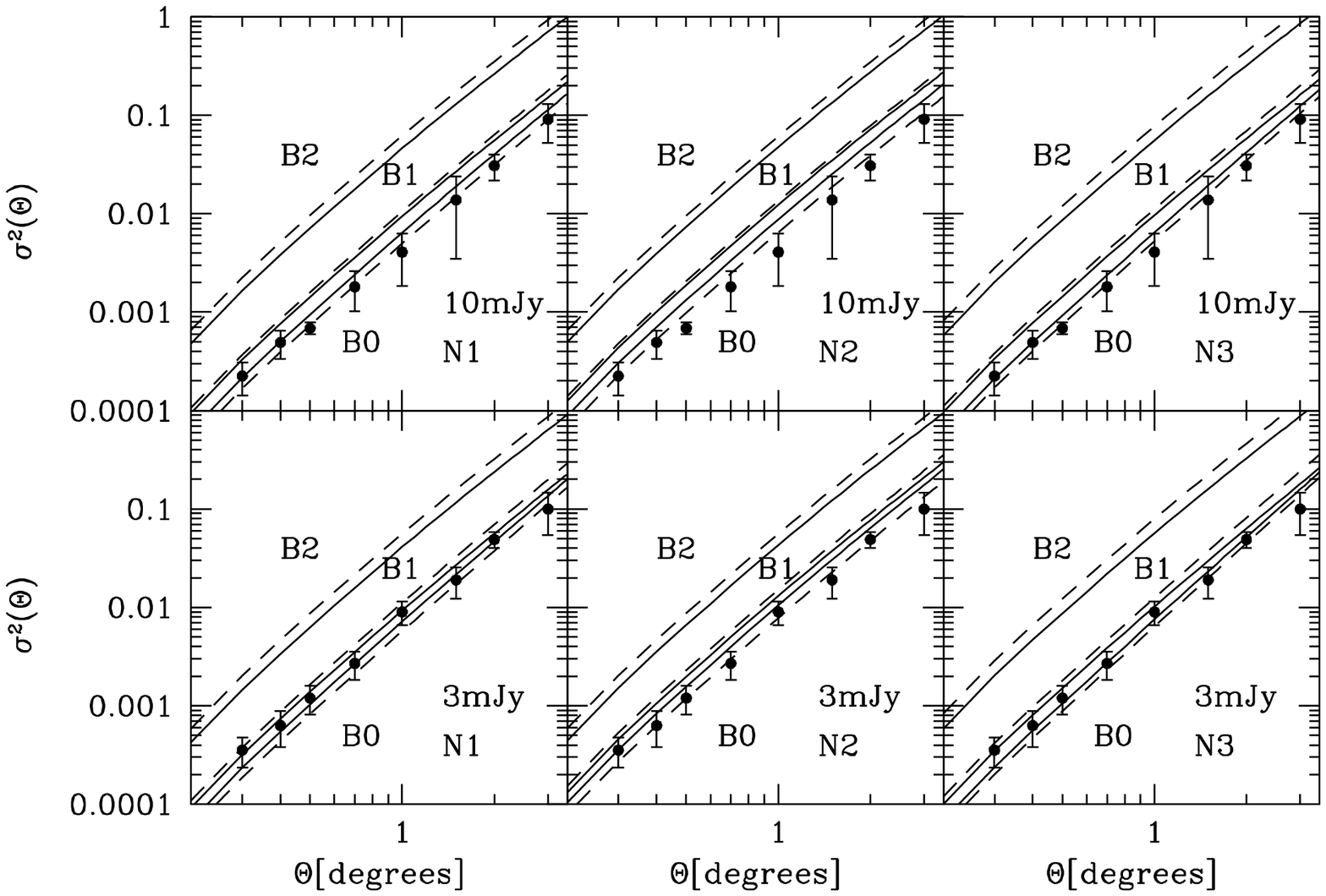} 
\caption{Theoretical predictions for the normalised variance $\sigma^2$ in
  the case of
  $\Gamma=0.2$, $N(z)\equiv N1,N2,N3$ at 10mJy and 3mJy, 
  and different bias models. The solid lines are for open models while the
  dashed lines are for spatially-flat models. The data are 
from the FIRST survey. 
\label{fig:fit1a} }
\end{figure*}   
 \begin{figure*}
\vspace{15cm}  % amount of vertical space needed
\includegraphics{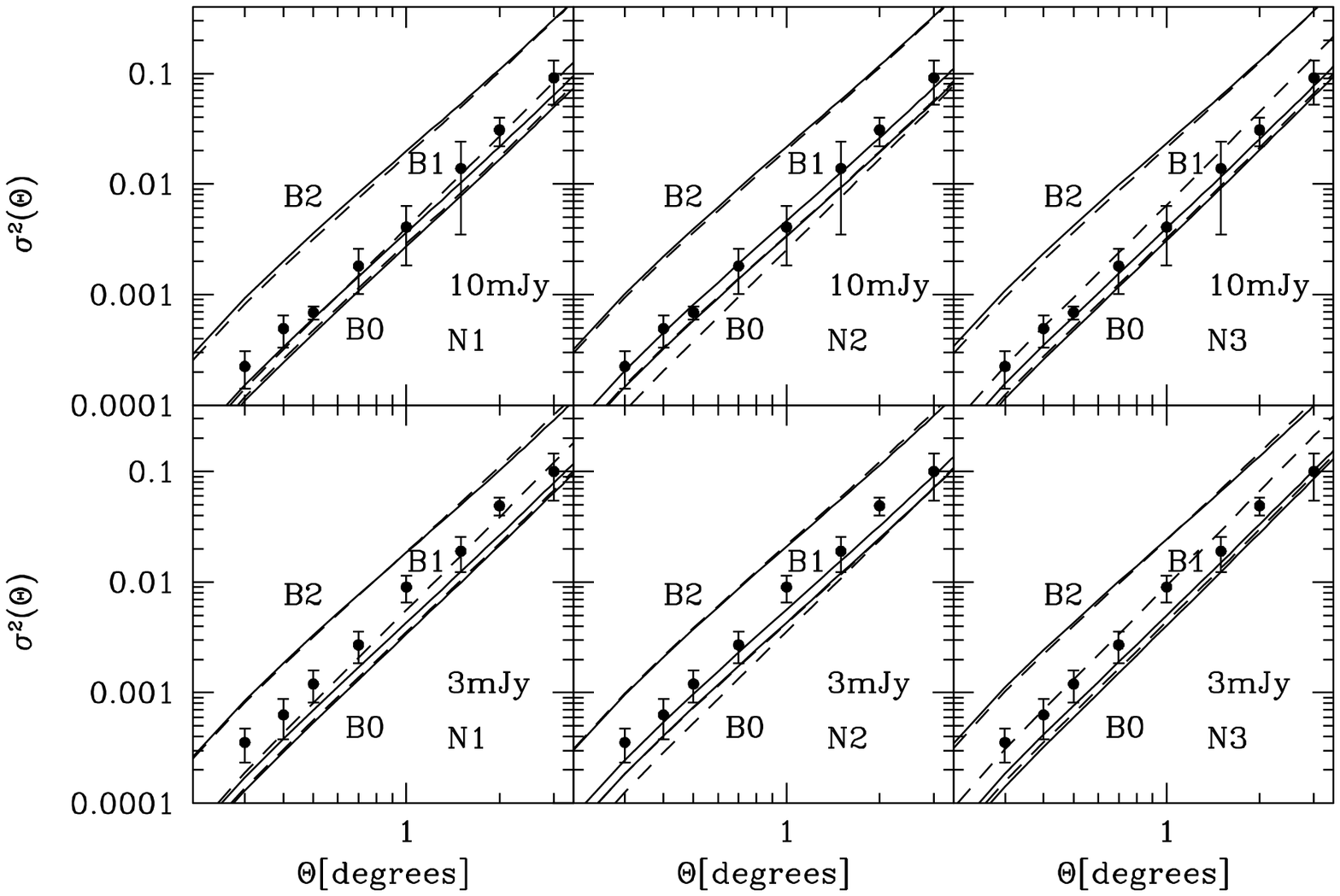}
\caption{Theoretical predictions for the normalised variance $\sigma^2$ in
  the case of
  $\Gamma=0.5$, $N(z)\equiv N1,N2,N3$ at 10mJy and 3mJy, 
  and different bias models. The solid lines are for open models while the
  dashed lines are for spatially-flat models. The data are 
from the FIRST survey. 
\label{fig:fit1b} }
\end{figure*}  

\noindent
From equations (\ref{eqn:sig2_fit5}),
(\ref{eqn:sig2_fit7}) and (\ref{eqn:sig2_fit10}) we have that the slope $\gamma$
of the correlation 
function (see eq. \ref{eqn:w_power}) is independent of the flux limit
in the range $3$mJy$\le F\le 10$mJy (see Magliocchetti et al., 1998
for the results at 3mJy). Furthermore we find that the
amplitudes $A$ at each flux-cut are vary smoothly with the flux limit, as
shown in figure \ref{fig:ampl}.
We can therefore be confident that the
clustering signal found in our analysis is mainly given by  
populations of radio objects which do not change their clustering 
properties as the flux level 
cut is increased. This result will be of
particular help in the next sections when we compare the theoretical
predictions for the correlation function with measurements 
at different flux limits.

\section {COMPARISONS WITH THE DATA}
One of the main reasons for preferring the variance statistic (i.e. the
counts in cells analysis) to the the direct evaluation of the
correlation function $\xi_{radio}$ is that $\sigma^2$ is 
a cumulative quantity which provides a
complete description of large-scale structure in a more robust way. 
Therefore, in order to match the models obtained for the angular
correlation function in section 4 with the results inferred from the
counts in cells analysis, one has to integrate
$w(\theta)$ to get the predictions for the variance of the
distribution of sources. 
This can be done by means of
equation (23) which, for a generic $w(\theta)$, is written as
\begin{eqnarray}
\sigma^2=\int w(\theta)\;\delta\Omega_1\;\delta\Omega_2,
\end{eqnarray}  
and in the case of square cells, 
can be expressed as the two dimensional integral
\begin{eqnarray}
\sigma^2(\Theta)=\Theta^2\int_0^{\Theta}dx\int_0^{\Theta}w(\theta)dy,
\end{eqnarray}
where $\Theta^2$ is the area of the cells and
$\theta=\sqrt{x^2+y^2}$. The quantity given by equation (29) has
then been evaluated for all the $w(\theta)$ models illustrated in
section 4 and the corresponding $\sigma^2$'s have been
compared with the estimates of $\sigma_{obs}^2$ obtained from the data
both at 3mJy and 10mJy as explained in section 5 and in Magliocchetti
et al. (1998).\\

\noindent
Figures 10 and 11 show the results for $\Gamma=0.2$ and
$\Gamma=0.5$ respectively; 
the symbols are as in section 4 and once again the
solid lines are for open models while the dashed lines are for
spatially-flat models. Independent of the functional form
adopted for the redshift distribution $N(z)$ and of the flux cut, 
models with bias strongly evolving with redshift can be
ruled by the data with a high
confidence level ($\simgt 3\sigma$) as they grossly overestimate the
amplitude of $\sigma^2(\Theta)$ at all scales.
Although the predictions for constant bias (B0) provide a somewhat better
fit to the data (particularly for $\Gamma = 0.2$), the B0 (constant bias) and
the B1 (bias linearly evolving with redshift) models are almost degenerate
with reference to the measurements.
%Although the predictions for 
%constant bias (B0) are slightly preferred for a fit to the
%data (this is especially true for $\Gamma=0.2$; in the case of
%  $\Gamma=0.5$ the situation is somehow different as all the models
%  are ``antibiased'' relative to the distribution of matter), 
%the B0 (constant bias) and the B1 
%(bias linearly evolving with redshift)
%models are almost degenerate in comparison with the
%measurements. 
As already mentioned in section 4, this effect is due to
the low normalisation of the mass correlation function
$\xi_{mass}(r)$ in the B1 case, resulting from the assumption that 
optical galaxies are
biased relative to the distribution of mass, and from the fact that for
B1 (contrary to the B2 models) the dependence of 
$b(z)$ on the redshift is not strong enough to drive the evolution of
$\xi_{radio}(r,z)$ to higher amplitudes, on these angular scales.\\
  If one assumes that radio sources are indeed
strongly biased with respect to ``normal'' galaxies by the factor
$\left(10/5.4\right)^{1.7}$ appearing in equations (15) and (16), then 
the conclusion of the B0 model as favoured for the interpretation of
the data is in agreement with the results found by 
Matarrese at al. (1997) and Frieman
\& Gazta\~naga (1994) from the angular correlation function in
optical/infrared surveys. \\

\noindent
This kind of analysis apparently is not sensitive to differences in
$N(z)$'s. As $\sigma^2$ is an integral quantity, its behaviour has
very little dependence on the form of the tail of $w(\theta)$ at large
angular scales where the clustering signal is weaker by at least two
orders of magnitude than that at $\theta\sim0.1^{\circ}$. Thus,
although the models for $w(\theta)$ for different redshift
distributions show remarkable differences, the combined effects of the
integral (28) and the presence of observational uncertainties in the
estimates of $\sigma_{obs}^2(\Theta)$ wash out these differences.

\section{THE CLUSTERING EVOLUTION FOR DIFFERENT RADIO POPULATIONS}
A major problem in the analysis and interpretation of the
clustering of radio sources at faint flux limits is the mixing of
different populations. The brighter objects 
in a deep radio survey are identified with powerful radio-loud
QSOs and AGN's; the fainter objects ($\simlt 3$mJy) are dominated by a
population of faint low-z sources which are local starbursting
galaxies (see Wall, 1994). 
As radio-loud QSOs tend to be hosted by
giant ellipticals within rich clusters (see e.g. Cristiani, 1998), 
these objects show a much stronger correlation
than normal field spirals (see also Loveday et al., 1995) containing
the population of
star-bursting galaxies. It is therefore of crucial importance
to take into account the characteristics of these two different
populations to get information on their clustering
properties and on the evolution of bias with redshift. 
Theoretical works suggest that bias depends on the mass of the dark
matter halos in which galaxies were born (Mo \& White, 1996) and
therefore its dependence can also be related to the mass of the galaxy
itself in the hypothesis that the most ``visible'' galaxies reside
within the most massive dark matter halos.\\

\noindent
To perform our analysis we use a simple
approach assuming at low redshifts the clustering signal is dominated
by the starbursting objects while at higher z's both the populations
contribute. This approximation is justified 
by the observational evidence of a drastic drop in the number of
quasars in the nearby universe. 
%The second assumption is probably not true, but leads to negligible
%errors, and  simplifies our analysis.
% while the second one comes from the
%fact that spirals and ellipticals choose very different
%environments and show very different clustering properties. In fact
%recent results (Steidel et al., 1998; Giavalisco et al., 1998)
%indicate that faint objects (in general field spirals), 
%supposed to be associated to
%less massive halos, are much less strongly clustered than bright
%galaxies residing in more massive halos giving rise to the 
%so-called phenomenon of {\it clustering segregation}. 
We therefore write the correlation function as follows:
\begin{eqnarray}
\xi_{tot}(r,z)=\left [A^2\xi_A\;  +2AB(\xi_A \xi_B)^{1/2}
+B^2\xi_B\;\right] 
%\left\{ \begin{array}{ll}
%\xi_{A}(r,z)=\xi(r) b'^2(z) D(z)^2 \;\;\; \mbox{$z<0.1$}\\
%A^2\xi_{A} (r,z)+B^2\xi_{B}(r,z)  \;\;\;\;\; \mbox{$z\ge 0.1$}
%\left [A^2\xi(r)b'^2(z) +2AB(\xi(r)\xi'(r))^{1/2} b(z)\;b'(z)\\
%+B^2\xi'(r) b^2(z)\right] D(z)^2 \;\;\;\mbox{$z\ge 0.1$}
%\end{array}
%\right .
\end{eqnarray}
where $\xi_A(r,z) $ is the correlation function for the starbursting
galaxies, $\xi_B(r,z)$ is for radio-loud quasars, and A and B are the
relative space densities the two populations normalized so that $A+B =
1$.
We assume that $ B = 0 $ for $z<0.1$, and consider three possibilities
for $z>0.1$: starburst dominated, $A=0.8, B=0.2$; an equal mix $A=0.5,
B=0.5$; and AGN dominated, $A=0.2, B=0.8$.
For the starburst correlation function we use $\xi_A(r,z)= \xi\; D(z)^2
b'(z)^2 $, where $\xi(r)$ and $D(z)$ are as expressed in section 3. 
For the AGN correlation function we use $\xi_B(r,z)= \xi'\; D(z)^2 b
(z)^2$ where $\xi'(r)$ is the effective linear bias mass correlation
function for QSOs: we assume on large scales that it is the linear
prediction for a CDM model with generic $\Gamma$, while at small scales 
$\xi$ has been extrapolated as a power-law of slope
$-2.1$ as found from the analysis of the clustering of bright-early
type galaxies in the Stromlo-APM redshift survey (Loveday et al.,
1995). Note that the value of 2.1 for the slope is different from that
used in section 3, where we considered the radio sample to be
homogeneous. $b(z)$ and $b'(z)$ are the bias associated with bright
radio sources (QSOs + AGN) and spiral galaxies (starbursting
population) relative to the distribution of mass; their expressions
are given by equations (15) and (16) assuming $r_0\sim 10h^{-1}$Mpc
(powerful radio sources) and $r_0\sim 5.4h^{-1}$Mpc
(optical-galaxies).
This yields for example, in the
case of constant bias (B0) models, that the starbursting population is
indeed unbiased ($b(z)=1$), while the AGN population shows a higher
bias relative to the distribution of the mass; at $z=0$,
$b(0)=\left(\frac{10}{5.4}\right)^{2.1/2}\frac{1}{\sigma_8^{mass}}$.  The
expression for $\xi_{tot}(r,z)$ has then been projected by means of
the Limber equation (1) in order to get the relative functional form
for the angular correlation function $w(\theta)$.\\ \begin{figure*}
\vspace{10cm}  % amount of vertical space needed
\includegraphics{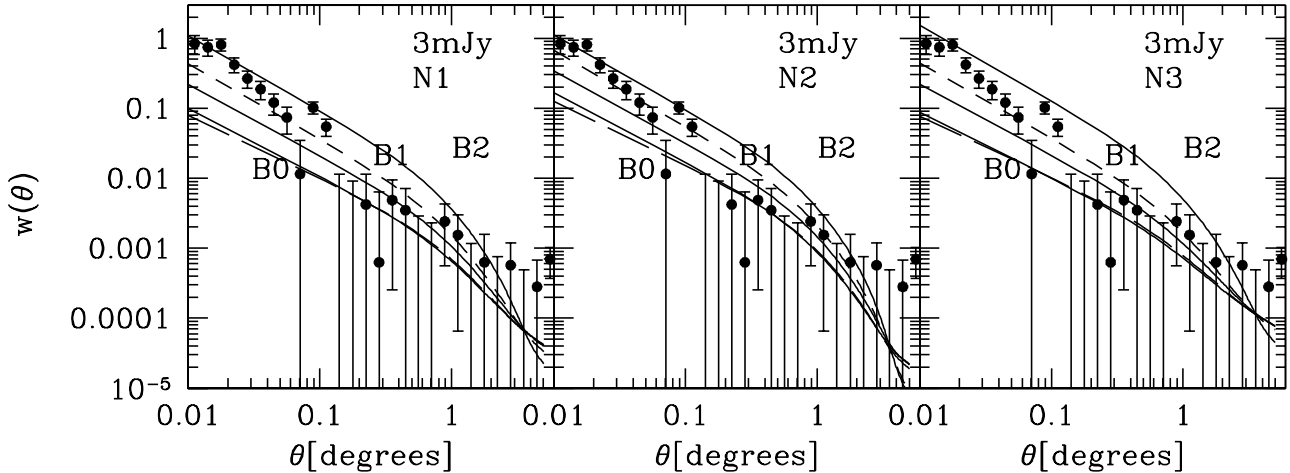} 
\caption{Theoretical prediction for the angular correlation function in
  the case of
  $\Gamma=0.2$, $N(z)\equiv N1,N2,N3$ at 3mJy, 
  and different bias models as
  shown in the figure. 
The solid lines are for models with $A=B=0.5$,
  the long-dashed lines are for $A=0.8$, $B=0.2$, while the short-dashed
  lines stand for $A=0.2$ and $B=0.8$ (see text for details). The data
  are obtained from measurements of $w_{obs}(\theta)$ using
equation~\ref{eqn:wdef}.  
The error bars show the Poisson estimates for the
  catalogue under consideration. 
\label{fig:fit1c} }
\end{figure*}   

\noindent
As Cress and Kamionkowsky (1998) have indicated, at scales
$\theta\simgt 1^{\circ}$ the angular correlation
function is
dominated by nearby objects (within $z\simlt 0.1$), while at scales 
$\theta\sim 0.1^{\circ}$ there is an equal contribution from local
($z\simlt 0.1$) and distant ($z\simgt 0.1$) sources. Thus, 
to study the evolution of the clustering of
bright distant objects we have to compare our models of
$w(\theta)$ with the observed correlation function
pushing the measurements down
to angles $\theta\sim 0.02^{\circ}$. Note that, as can be seen in
figure 4, the assumption of growth of clustering under linear theory
is still a good approximation at these angular scales. 
The counts-in-cells statistics for the FIRST survey are dominated by
shot (Poisson) noise for $\theta \simlt 0.3^{\circ}$ (see
Magliocchetti et al., 1998) and so to extend the range of measurement
down to smaller scales we have measured angular correlation function
$w_{obs}(\theta)$ directly from the catalogue. 

\noindent
We recall here that the 
angular correlation function $w_{12}=w(\theta)$ gives the excess of
probability, with respect to a Poisson random distribution, of finding
two sources in the solid angles $\delta\Omega_1$ $\delta \Omega_2$
separated  by an angle $\theta$, and it is defined as
\begin{eqnarray}
\delta P=n^2 \delta\Omega_1\delta \Omega_2  \left[1+w(\theta)\right]
\end{eqnarray}
where $n$ is the mean number density of objects in the catalogue under
consideration. We measured $w_{obs}$ from the estimator (Hamilton, 1993)
\begin{eqnarray}
w_{obs}(\theta)=4 \frac{DD\;\;RR}{DR^2}\;\;-1
\label{eqn:wdef}
\end{eqnarray}
where $DD$ is the number of distinct data-data pairs, $DR$ is the
number of random-data pairs and $RR$ is the number of random-random
pairs as a function of angular separation and where the random data 
have been obtained by generating a random
catalogue of positions and selected sources above the sensitivity
limit from the coverage map (see Magliocchetti et al., 1998 for further
information about the catalogue).\\

\noindent
Given the very weak dependence of the angular 
correlation function
on geometry, the shape parameter
of the power spectrum $\Gamma$ and the flux cut (at least for
$F\le10$mJy), we restrict our analysis to
$\Omega=1$, $\Gamma=0.2$ and $F\ge 3$mJy, while still allowing for different
functional forms of the redshift distribution $N(z)$. 
All the models assume that the
distribution of starburst objects is unbiased with respect
to the mass distribution, justified by the results obtained for
the more local population in section
6. The corresponding models are shown in figure 12 where we also
plot $w_{obs}$ taken from the FIRST survey at 3mJy; the error bars 
show Poisson estimates for the catalogue under consideration. 
The solid lines are for
models in which the distribution of the AGN population is 
unbiased (B0), has a bias linearly evolving with
redshift (B1) and has a bias strongly evolving with z (B2). The
predictions have been calculated in the case of the two populations
equally contributing to the clustering signal
($A=B=0.5$). As figure 12 shows, while at large angular scales the B0 and
B1 models are roughly degenerate as in section 6, pushing the
analysis down to smaller scales removes this degeneracy. In fact 
the B0 and B2 models are ruled out as possible descriptions
of the data as they show either too small (B0) or too
large (B2) amplitude, especially at intermediate scales. 
The best fit is provided by the B1
curves i.e. by models in which the distribution of the starbursting 
population is taken to be roughly unbiased while that one for 
the AGN population shows a bias linearly evolving with redshift.

The effect of varying the relative weight of the two populations in
the B1 model is shown in figure 12 for the two cases: $\sim 20\%$
starburst and $\sim 80\%$ AGN (short dashed lines) and $\sim 80\%$
starburst and $\sim 20\%$ AGN (long dashed lines). The agreement with
the data is best for the AGN dominated model.
%in the case the N2 redshift
%distribution (dominated by a 
%bump at $z\sim 1$).
% as opposed to the descriptions provided by
%N1 and N3, pointing out the underestimate of
%the number of bright objects at high redshifts in both the N1 and N3 models.
%Given the errors in the estimates of 
%$w_{obs}(\theta)$ at large angular scales, one has that it is
%otherwise impossible to discern amongst
%models obtained for different $N(z)$'s as all the remarking
The notable differences for different $N(z)$'s found in section 4 
reside primarily in the part of the angular correlation function
at large scales where the signal is dominated by the noise.\\
\\ 
The fact that we need a biased distribution of AGN in order to match
the data is in agreement with the hierarchical models of galaxy
formation predicting the less massive objects to be less
strongly clustered than those formed in the higher and biased 
peaks of the density
field (Giavalisco et al., 1998), with the most powerful sources being
hosted by the most massive dark matter halos, and the biasing factor
depending (at least) both on the epoch and the type of objects
under consideration (i.e. on the halo mass).
%the more massive objects to be more clustered than
%the less massive ones, with the more massive (and bright) objects forming
%from the higher and biased peaks of the density field. 
%As a further test of this, we varied in the B1 model the
%relative weight of two populations, namely by assuming
%a $\sim 70\%$ starburst and $\sim 30\%$ AGN (long
%dashed lines in figure 14) or the opposite ratio (short
%dashed lines in figure 14). The 
%resulting models do not differ very much; nevertheless
%it is worth noting that the best fit to the data is given by
%those curves worked out on the assumption of the starburst 
%population being more abundant than the AGN population. 
%Once again this could be very easily explained by a
%hierarchical model in which the more abundant and faint objects are less
%strongly clustered than those residing in relatively rare peaks of the density
%field (Giavalisco et al., 1998), with the most powerful sources being
%hosted by the most massive dark matter halos, and the biasing factor
%depending (at least) both on the epoch and the type of objects
%under consideration (i.e. on the halo mass).  

\section{CONCLUSIONS}
To put constraints on the redshift
distribution $N(z)$ of radio objects at faint fluxes and on their
clustering and biasing properties, we have calculated 
predictions for the angular correlation
function $w(\theta)$ using three models for $N(z)$ and three models for the
clustering evolution. The three models for the redshift distribution have 
been chosen to span the range of reasonable
possibilities.  The predictions for $w(\theta)$ have 
been calculated for a $\xi_{mass}$ obtained from two CDM
models ($\Gamma=0.2$ and $\Gamma=0.5$), three different
assumptions for the evolution of the bias and in the case
$\Omega_0=0.4$, 
both flat and open space. We chose a fixed value for
the density parameter because of the degeneracy amongst predictions
obtained for different $\Omega_0$'s. The bias models
correspond to three sensible galaxy evolution models (galaxies tracing
the mass at all epochs - constant bias; $b(z)\propto(1+z)$ -
galaxy conserving model; $b(z)\propto(1+z)^{1.8}$ - galaxy merging
model).\\ 

\noindent
The discrimination amongst different models comes both from the fall
of the calculated $w(\theta$)
at large angular scales and from its overall amplitude. 
The fall depends on the
negative feature in $\xi$ at large scales, which is a generic 
feature of CDM-like models, and also on the low-z component in the
N(z) whose amplitude depends on
the presence and relative contribution of any local population 
of radio objects. The overall
amplitude of the angular correlation function is instead mainly
related to the level of bias at high redshift.\\

\noindent
For a comparison with the data inferred from the counts-in-cells
analysis of the FIRST survey we integrated the models for
$w(\theta)$ to calculate the corresponding predictions for the
variance of the distribution $\sigma^2$; we then matched the results
with the measurements for different flux limits presented in section 1
and in Magliocchetti et al. (1998). Although models for
the angular correlation function obtained for different $N(z)$ show
quite remarkable differences, in the framework of this analysis it is
not possible to discriminate amongst them, given the errors in the
measurements and the fact that the integral used to calculate
$\sigma^2$ washes out the differences.\\
Concerning the bias evolution, all the models of bias strongly 
dependent on redshift can be ruled out with
high confidence level as they grossly overestimate the amplitude of
$\sigma^2$ at all scales. There is a degeneracy between models with
constant bias and those for bias depending
linearly on redshift, although the predictions for constant bias 
are mildly preferred by the data, especially for
$\Gamma=0.2$. The analysis is quite insensitive to the geometry of the 
Universe.\\

\noindent
As a further step, we introduced a model for the correlation function 
able to account for a mixing of populations with completely 
different properties,
the first population of faint starbursting/spiral galaxies, the
second one of AGN-powered objects normally hosted by bright
ellipticals. We assumed that
starbursting population traces the distribution of mass at all
times, and we allowed for biasing evolution in the
case of radio-loud AGN. Appropriate measurements for 
$w_{obs}(\theta)$ were made directly from the catalogue. 
It turns out that an evolution of
bias with redshift is required for AGN to match the data. The
model that best fits the observed angular correlation function is
given by the sum of a) a less numerous population of faint starbursting
sources, with clustering properties very similar to normal spiral
galaxies and a distribution that is roughly unbiased relative to the
distribution of mass, and b) a more numerous population of 
radio-loud AGN's much more strongly clustered and whose
distribution is biased relative to the distribution of the mass,
with a bias factor which evolves linearly with redshift.
This result is in good agreement with hierarchical CDM theories of galaxy
formation that predict the lower luminosity/lower mass objects being less
clustered than the high-luminosity sources, with the brighter sources
being associated with higher (and biased) peaks in the dark matter
distribution.\\
           
\noindent
There are crucial observations needed to further the
analysis: in particular observational measurements of N(z), from
identifications and redshift measurements for a complete sample, would
be of great importance in constraining structure formation models as
well as understanding the distribution and
evolution with time of the
different populations of radio objects at mJy levels.

\vspace{1cm}
\noindent
{\bf ACKNOWLEDGEMENTS}\\
MM acknowledges support from the Isaac Newton Scholarship. We thank
George Efstathiou, Jarle Brinchmann and Jasjeet Bagla for helpful discussion.


\begin{thebibliography}{}
\bibitem[\protect\citename{Bagla}1997]{Bag}
Bagla, J.S., 1997;
astro-ph/9711081
\bibitem[\protect\citename{Baugh et al. }1996]{Ba}
Baugh C.M., Cole S., Frenk C.S., 1996;
{\it MNRAS}, {\bf282}, L27
\bibitem[\protect\citename{Bond}1984]{Bo}
Bond J.R., Efstathiou G., 1984; {\it ApJ}, {\bf 285}, L45
\bibitem[\protect\citename{Becker et al. }1995]{Be}
Becker R.H., White R.L., Helfand D.J., 1995; {\it ApJ}, {\bf 450}, 559
\bibitem[\protect\citename{Broadhurst et al. }1992]{Broa}
Broadhurst T.J., Ellis R.S., Glazebrook K., 1992;
{\it Nat}, {\bf355}, 55
\bibitem[\protect\citename{Bunn et al. }1997]{Bunn}
Bunn E.F., White M., 1997; {\it ApJ}, {\bf 480}, 6
\bibitem[\protect\citename{Clements }1996]{Cle}
Clements D.L., Couch W.J., 1996;
{\it MNRAS}, {\bf280}, L43
\bibitem[\protect\citename{Cress et al. }1996]{Cr}
Cress C.M., Helfand D.J., Becker R.H., Gregg M.D., White R.L., 1996; 
{\it ApJ}, {\bf 473}, 7
\bibitem[\protect\citename{Cress }1998]{Cres2}
Cress C.M., Kamionkowsky M., 1998;
astro-ph/9801284
\bibitem[\protect\citename{Cristiani }1998]{Cristiani}
Cristiani S., 1998; 1998elss.confE..41C
\bibitem[\protect\citename{Dunlop }1990]{Dun}
Dunlop J.S., Peacock J.A., 1990;
{\it MNRAS}, {\bf247}, 19
%\bibitem[\protect\citename{Efstathiou }1991]{Ef}
%Efstathiou G., Bernstein G., Katz N., Tyson J.A., Guhathakurta P.;
%{\it ApJ}, {\bf 380}, L47
\bibitem[\protect\citename{Frieman \& Gazta\~naga }1994]{Fri}
Frieman J.A., Gazta\~naga E., 1994; {\it ApJ}, {\bf 425}, 392
\bibitem[\protect\citename{Fry }1996]{Fry}
Fry J.N. 1996;
{\it ApJ}, {\bf 461}, L65
\bibitem[\protect\citename{Giavalisco }1998b]{Gia}
Giavalisco M. et al. 1998b; in preparation
\bibitem[\protect\citename{Hamilton }1993]{Ha1}
Hamilton A.J.S., 1993; {\it ApJ}, {\bf 417}, 19
\bibitem[\protect\citename{Hamilton et al. }1991]{Ha}
Hamilton A.J.S., Kumar P., Lu E., Matthews A. 1991; {\it ApJ}, {\bf 374}, L1
\bibitem[\protect\citename{Kooiman et al. }1995]{Ko}
Kooiman L.K., Burns J.O., Klypin A.A., 1995; {\it ApJ}, {\bf 448}, 500
\bibitem[\protect\citename{Kundic et al. }1997]{Ku}
Kundic T., Turner E.L., Colley W.N., Gott J.R.III,
Rhoads J.E., Wang Y., Bergeron L.E., Gloria K.A., Long D.C., Malhotra
S., Wambsganss J., 1997; {\it ApJ}, {\bf 482}, 75
\bibitem[\protect\citename{Lahav et al. }1991]{La1}
Lahav O., Lilje P.B., Primack J.R., Rees M.J., 1991;
{\it MNRAS}, {\bf 251}, 128
\bibitem[\protect\citename{Lahav et al. }1992]{La}
Lahav O., Saslaw W.C., 1992; {\it ApJ}, {\bf 396}, 430
\bibitem[\protect\citename{Loan et al. }1997]{Lo}
Loan A.J., Wall J.V., Lahav O., 1997; {\it MNRAS}, {\bf 286}, 994
\bibitem[\protect\citename{Loveday et al. }1995]{Lov}
Loveday J., Maddox S.J., Egstathiou G., Peterson B.A.; 1995
{\it ApJ}, {\bf 442}, 457
\bibitem[\protect\citename{Maddox et al. }1990]{Ma}
Maddox S.J., Efstathiou G., Sutherland W.J., Loveday J., 1990;
{\it MNRAS}, {\bf242}, 43P
\bibitem[\protect\citename{Magliocchetti et al. }1998]{Mag}
Magliocchetti M., Maddox S.J., Lahav O., Wall J.V., 1998;
astro-ph/9802269
\bibitem[\protect\citename{Matarrese et al. }1997]{Mat}
Matarrese S., Coles P., Lucchin F., Moscardini L., 1997;
{\it MNRAS}, {\bf286}, 115
\bibitem[\protect\citename{M0 et al. }1996]{Moo}
Mo H.J., White S.D.M., 1996;
{\it MNRAS}, {\bf282}, 347
\bibitem[\protect\citename{Moscardini et al. }1997]{Mo}
Moscardini L., Coles P., Lucchin F., Matarrese S., 1997;
astro-ph/9712184
\bibitem[\protect\citename{Pea }1991]{Pea2}
Peacock J.A., Nicholson D., 1991;
{\it MNRAS}, {\bf253}, 307
\bibitem[\protect\citename{Peacock }1994]{Pea}
Peacock J.A., Dodds S.J. 1994;
{\it MNRAS}, {\bf267}, 1020
\bibitem[\protect\citename{Peacock }199]{Pea1}
Peacock J.A., 1997;
{\it MNRAS}, {\bf 284}, 885
\bibitem[\protect\citename{Peebles }1980]{Pe}
Peebles P.J.E., 1980; {\it The Large-Scale Structure of the Universe},
Princeton University Press
\bibitem[\protect\citename{Peebles1 }1984]{Pe1}
Peebles P.J.E., 1984; {\it ApJ}, {\bf 284}, 439
\bibitem[\protect\citename{Seldner }1981]{Se}
Seldner M., Peebles P.J.E., 1981; {\it MNRAS}, {\bf 194}, 251
\bibitem[\protect\citename{Shaver et al. }1989]{Sha}
Shaver P.A., Pierre M., 1989; {\it A\&A}, {\bf 220}, 35
\bibitem[\protect\citename{Sicotte }1995]{Si}
Sicotte H, 1995; {\it PhD thesis}, Princeton University
\bibitem[\protect\citename{Steidel et al. }1996]{Ste1}
Steidel C.C., Giavalisco M., Pettini M., Dickinson M., Adelberger
K.L., 1996; {\it ApJ}, {\bf 462}, L17
\bibitem[\protect\citename{Steidel et al. }199]{Ste2}
Steidel C.C., Adelberger K.L., Dickinson M., Giavalisco M., Pettini
M., Kellog M., 1997; atro-ph/9708125
\bibitem[\protect\citename{Steidel et al.}1998]{Ste3}
Steidel C.C., Adelberger K.L., Giavalisco M., Dickinson M., Pettini
M., Kellog M., 1998; atro-ph/9805267
\bibitem[\protect\citemname{Tegmark et al.}1998]{Te}
Tegmark M., Peebles P.J.E., 1998; astro-ph/980407
\bibitem[\protect\citemname{Totsuji et al.}1969]{To}
Totsuji H., Kihara T., 1969; {\it PASJ}, {\bf 21}, 221
\bibitem[\protect\citename{Treyer }1996]{Tre}
Treyer M.A., Lahav O. 1996;
{\it MNRAS}, {\bf 280}, 469
\bibitem[\protect\citename{Tyson }1988]{Ty}
Tyson J.A., 1988;
{\it AJ}, {\bf 91}, 1 
\bibitem[\protect\citename{Wall}1994]{Wa} 
Wall J.V., 1994; {\it Austr. J. Phys.}, {\bf 47}, 625
\bibitem[\protect\citename{Wall, Jackson }1997]{Wa1}
Wall J.V., Jackson C.A., 1997;
{\it MNRAS}, {\bf 290} L17
\bibitem[\protect\citename{Windhorst et al.}1985]{Wi} 
Windhorst R.A., Miley G.K., Owen F.N., Kron R.G., Koo R.C., 1985; 
{\it ApJ}, {\bf 289}, 494
\end{thebibliography}
\end{document}